\documentclass[12pt,letterpaper]{article}

\usepackage{amsmath,amssymb,calc}
\usepackage{graphicx}
\usepackage{color}

\newcommand\be{\begin{equation}}
\newcommand\ee{\end{equation}}
\newcommand{\bea}{\begin{eqnarray}}
\newcommand{\eea}{\end{eqnarray}}

\newcommand{\nn}{\nonumber}
\newcommand{\pd}{\partial}

\def\id{\protect{{1 \kern-.28em {\rm l}}}}

\def\1{^{(1)}}
\def\0{^{(0)}}
\def\2{^{(2)}}

\def\id{\protect{{1 \kern-.28em {\rm l}}}}

\let\non\nonumber

\begin{document}
\begin{titlepage}
\begin{center}
\hfill \\

\vskip 1cm 
{\Large \bf Stability of D-brane embeddings in \vspace{0.3cm}\\nontrivial backgrounds}
\vskip 1.5 cm
{\bf  Lilia Anguelova${}^{a,c}$\footnote{languelova@perimeterinstitute.ca}, Peter Suranyi${}^b$\footnote{peter.suranyi@gmail.com} and L.C.R. Wijewardhana${}^b$\footnote{rohana.wijewardhana@gmail.com}\\
\vskip 0.5cm  {\it ${}^a$ Perimeter Institute for Theoretical Physics, Waterloo, ON N2L 2Y5, Canada\\ ${}^b$ Department of Physics, University of Cincinnati,
Cincinnati, OH 45221, USA\\ ${}^c$ Institute for Nuclear Research and Nuclear Energy, BAS, Sofia, Bulgaria} \non\\}

\end{center}
\vskip 2 cm
\begin{abstract}
We propose a new analytical method for determining whether nonsupersymmetric probe branes embedded in nontrivial backgrounds are perturbatively stable or not. The method is based on a relationship between zero mass solutions of the relevant DBI equations of motion and tachyonic solutions. Furthermore, due to the above relation, the question, of whether a classical solution is stable or not, can be answered simply by studying the derivatives of that solution with respect to its integration constants. Finally, we illustrate the efficiency of this method by applying it to several interesting examples.
\end{abstract}
\end{titlepage}

\tableofcontents

\section{Introduction}

Holographic methods \cite{JM}-\cite{GGDmore_gen} are a powerful tool for addressing
hitherto unsolvable problems in field theory by connecting strongly coupled gauge theories with weakly coupled gravity backgrounds in higher dimensions. These methods rely on embedding probe branes of various dimensions in appropriate graviational and flux backgrounds. The shape of those branes is determined, at the classical level, by extremizing the Dirac-Born-Infeld (DBI) action that describes them. In many examples, the probe branes are supersymmetric and therefore the classical shape of the embedding is guaranteed to be stable under small fluctuations. However, many other important applications (see for example \cite{KS}-\cite{CLtV}), which are of great phenomenological interest, require the study of nonsupersymmetric probes. Hence one needs to compute explicitly the mass spectrum of fluctuations, in order to see whether there is a tachyonic mode, signifying perturbative instability, or not.

The computation of the mass spectrum is performed in the following way. First, one finds the linearized equations of motion for small fluctuations around the classical solution of interest. To solve those equations, one then usually uses an ansatz, which factorizes into a multiplier dependent on the 4d space-time coordinates and another multiplier dependent on a radial coordinate of the background metric. The 4d space-time dependent factor admits a plane wave ansatz. Then, after performing a certain field redefinition, one can reduce the problem to solving a Schrodinger-like equation in the radial direction, with the role of the energy eigenvalue being played by the square of the mass $m$ characterizing the flucutation (for an example of this procedure, see \cite{ASW2}). Solutions with $m^2>0$ belong to the physical mass spectrum of the theory, whereas the presence of modes with $m^2<0$ indicates perturbative instability.

Usually, it is rather difficult to solve the appropriate Schrodinger equation analytically; it is only possible to study it with numerical techniques. To facilitate the investigation of the question of stability, although not the problem of finding the full mass spectrum, we develop a method relying entirely on the investigation of $m=0$ solutions of the Schrodinger equation. Often, finding such solutions is much simpler than solving the complete mass eigenvalue problem. In fact, as we will point out below, once a classical solution has been found, obtaining the $m=0$ fluctuations around it is straightforward by analytical means. The existence of zeros of those fluctuations, as functions of the radial variable, then indicates perturbative instability. 

In the next section we will show that zero mass fluctuations around classical solutions do encode the information about whether there are tachyonic modes in the spectrum of shape fluctuations or not. This is an expanded explanation, based on the more concise arguments presented in Appendix B of \cite{ASW3}. In Section 3, we show that the analytic form of zero mass fluctuations can be obtained directly from the classical solution under consideration, by differentiating it with respect to its integration constants. In Section 4, we illustrate our method by applying it to three models that are of phenomenological interest: the Kuperstein-Sonnenschein model of chiral symmetry breaking \cite{KS}, the Sakai-Sugimoto construction of holographic QCD \cite{SS} and the walking technicolor model of \cite{LA,ASW}. In the last section, we summarize our results.

\section{Stability from fluctuations with vanishing mass}

We begin by briefly reviewing the basic set-up for the models of interest. To build a gravitational dual of a strongly coupled gauge theory, one considers a background sourced by a stack of D$p$ branes. This gives rise to the color degrees of freedom. Adding flavor ones is achieved usually by introducing some D$q$ probe branes into that background.\footnote{Taking into account the full backreaction of the flavor branes is a very hard technical problem. To date, there is only rather modest progress toward solving it, in rare cases and/or with specific approximations. We will not discuss this issue further here.} The ten-dimensional metric describing a stack of $N_c$ coincident D$p$ branes, with $p=0,1,...,6$\,, is of the form \cite{HS}:
\be \label{HS}
ds_{10d}^2 = H^{-1/2} (y) \,dx^2_{(p+1)} \,+ \,H^{1/2} (y) \,dy^2_{(9-p)},
\ee
where $x_0,...,x_p$ are the $p+1$ worldvolume coordinates and $dx^2_{(p+1)} \equiv - dx_0^2 + ... + dx_p^2$\,, while the $y_1,...,y_{9-p}$ are the coordinates of the $10 - (p+1)$ transverse directions and $dy^2_{(9-p)} \equiv dy_1^2 + ... + dy_{9-p}^2$. Also, $H(y)$ is a harmonic function on the transverse space, containing the parameter $N_c$. Note that one can trivially write the transverse space line element in spherical coordinates:
\be
dy^2_{(9-p)} = dr^2 + r^2 d\Omega^2_{8-p} \, .
\ee
To obtain the dual of an effective lower (than $p+1$) dimensional gauge theory, one first needs to take a certain decoupling limit \cite{JM,GGDmore_gen} and then wrap the D$p$ branes on a compact manifold of appropriate dimension. For example, to obtain a four-dimensional gauge theory from a stack of D4 branes, one has to wrap them on an $S^1$. Similarly, when considering D5 branes, one needs to compactify them on an $S^2$, in order to end up, at low energies, with an effective four-dimensional description. These two steps, namely performing the decoupling limit and the subsequent compactification, lead to a metric of the form:
\be \label{bkgrM}
ds^2 = f_1 (U) \,dx^2_{(k)} + f_2 (U) \,dU^2 + ds^2_{9-k} (U) \, ,
\ee
where $U$ is $r$ up to a constant factor, $dx^2_{(k)}$ is the line element of the effective $k$-dimensional theory we want to study and $f_{1,2} (U)$ are functions that differ from case to case.  
For many examples, including the Sakai-Sugimoto model \cite{SS}, the line element $ds^2_{9-k} (U)$ is of the form:
\be \label{SSmlp}
ds^2_{9-k} = f_3 (U) \,dx^2_{(p+1-k)} + f_4 (U) \,d \Omega^2_{8-p} \, ,
\ee
where $dx^2_{(p+1-k)}$ corresponds to the compactified worldvolume dimensions and the functions $f_{3,4} (U)$ are case specific. There are more involved examples, though, in which one needs to perform certain 'twisting' while compactifying. This leads to a more complicated expression for $ds^2_{9-k}$, containing mixing between the two terms of (\ref{SSmlp}) as well as additional warp factors; see equation (6) of \cite{NPP}, for instance. However, for later purposes, it is important to underline that in all cases the background metric {\it does not} have a mixed component between the $U$ coordinate and the coordinates $\{x_i\}|_{i=0,...,k-1}$. This property is inherited from (\ref{HS}) and, more precisely, from the lack of a mixed $dx dr$ term there. 

As already mentioned, the gauge theory of interest lives in the $(x_0,...,x_{k-1})$ spacetime dimensions of the above background. To add flavor degrees of freedom in this theory, we embed some D$q$ probe branes in the background. Those probes are described by the usual DBI action: 
\be
S_{DBI} = -T \int d^{\,q+1} \sigma \,e^{-\phi} \sqrt{-\det (g_{ab} + 2\pi \alpha' F_{ab})} \, ,
\ee
where $a,b=0,...,q$ are worldvolume indices, $T$ is the brane tension, $\phi$ is the string dilaton, $g_{ab}$ is the metric induced on the worldvolume and $F_{ab}$ is the worldvolume field strength. For the class of holographic models we consider, the probe branes always extend along $x_0,...,x_{k-1},U$ and some number of compact directions. Integrating out that compact space, one is left with an effective action of the form:
\be
S = \int d^k x \,dU \,{\cal L} \, .
\ee 
Since we are primarily interested in four-dimensional effective theories, we will focus on the $k=4$ case from now on, although our considerations easily generalize to $k\neq 4$. Also, to underline the fact that in many examples the most suitable worldvolume radial variable may be a nonlinear function of the original spacetime radial variable, we introduce a new radial variable $z = z (U)$ and therefore the effective action acquires the schematic form:
\be
S = \int d^4 x \,dz \,{\cal \hat{L}} \, .
\ee 
Another important assumption is the following. The world volume field strength $F_{ab}$ does not have a nontrivial background. In other words, the only contribution to it comes from fluctuating its potential. This is certainly the case for the D8 probe branes in the Sakai-Sugimoto model \cite{SS} and the D7 probes in the Kuperstein-Sonnenschein model \cite{KS}, as well as for the probes in all the literature on holographic models of technicolor; see \cite{CES}-\cite{ASW2}, for example. The motivation for this assumption is that a generic nonvanishing background for $F_{ab}$ would lead to some kind of a monopole background in the four-dimensional effective theory. And this is not what the above models aimed to study. Nevertheless, it may be possible to consider a nontrivial $F_{ab}$ background that extends only along the compactified D$q$ directions, as well as possibly $z$. In fact, it may be that such a worldvolume flux is needed in order to stabilize certain embeddings, as mentioned in \cite{ASW3}. We leave the investigation of nonvanishing worldvolume flux for the future. Instead, our goal here will be to study the stability of probe brane embeddings under fluctuations of their shape (and with no worldvolume flux included). Therefore, the effective action of interest for us is of the form:
\be \label{Dir}
S = const \int d^4 x\,dz \,\sqrt{-\det g_{\hat{a}\hat{b}}} \, ,
\ee
where $\hat{a},\hat{b}=x_0,...,x_3,z$ \,. 

Let us denote by $\psi_1,...,\psi_{9-q}$ the coordinates transverse to the probe brane worldvolume. For simplicity, we will consider only one such coordinate $\psi$ in the following. To describe the embedding of the worldvolume into the background spacetime (\ref{bkgrM}), we need an ansatz for the dependence of $\psi$ on the worldvolume coordinates. As in all of the existing literature in this area (see, for example, \cite{KS,SS}, as well as \cite{CES}-\cite{MS}), we will assume that the classical embedding function depends only on $z$, but not on $x^{\mu}$. Its form is then easily determined by solving the equation of motion of (\ref{Dir}). The fluctuations $\delta \psi (z,x^{\mu})$ around that classical shape $\psi_{cl} (z)$ give rise to the scalar mesons in these models. Our goal will be to develop an efficient short-hand method for establishing when the spectrum of $\delta \psi (z,x^{\mu})$ has a negative mass-squared mode and thus implies that the original embedding is not stable. Note that, had we allowed the classical solution to be a function of $x^{\mu}$ as well, then we would not have had a standard four-dimensional Lagrangian (in $x^{\mu}$ spacetime) for the fluctuations as it would have depended on $x^{\mu}$ explicitly (as a result of substituting the expression for the classical solution) and not just implicitly via the fluctuation fields.\footnote{Nevertheless, it might be interesting, for some other applications, to explore classical embeddings that depend on $x^{\mu}$. 
Such a possibility arises when one imposes $x^{\mu}$-dependent boundary conditions.} Let us also remark that if there are several fields $\psi$ (and thus several $\delta \psi$'s), but no mixed terms to second order in their fluctuations, one gets several copies of the considerations we will develop here for a single field. 
This is the case in all of the examples in \cite{KS} and \cite{ASW}-\cite{CLtV}, whereas in \cite{SS} there is only one transverse coordinate. In principle, though, it is possible to have mixed terms between different $\delta \psi$ fields, even to second order, for more complicated embeddings. It would be interesting to see whether the presence of such terms could lead to anything new. We leave that question for the future.

To obtain the equation of motion for $\delta \psi (z,x^{\mu})$, one first substitutes the decomposition $\psi = \psi_{cl} (z) + \delta \psi (z,x^{\mu})$ into (\ref{Dir}). The result is an action of the form:
\be
S = \int dx d^4 x \sqrt{F_1 [\psi (z,x^{\mu}), \pd_z \psi (z,x^{\mu}), z] + F_2 [\psi (z,x^{\mu}),z] \,[\pd_{\nu} \psi (z,x^{\mu}) \pd^{\nu} \psi (z,x^{\mu})]} \, ,
\ee
where we have taken into account that, as usual in this area, we will work only to second order in fluctuations and also that the background metric does not have a non-vanishing $(z\mu)$ component, as explained above. This form of the action will be enough for our needs in Section 3 below. However, before that, it will be very useful to make connection with the previous literature, in which a Schrodinger form of the field equation was discussed. For that purpose, let us specify a bit more the form of the action that one obtains from (\ref{Dir}), still keeping only up to second order in fluctuations under the square root. Namely, we have:
\be \label{A3Fs}
S = \!\!\int \!\!dz d^4 x \sqrt{F_1^{(1)} (\psi_{cl} \!+ \!\delta \psi , z) +F_1^{(2)} (\psi_{cl} \!+ \!\delta \psi , z) \,[\psi'_{cl} \!+ \!\delta \psi']^2 + F_2 (\psi_{cl} \!+ \!\delta \psi , z) \pd_{\mu} \delta \psi \pd^{\mu} \delta \psi} \, ,
\ee
where for convenience we have introduced the notation $'\equiv \pd_z$. The Euler-Lagrange equation for $\delta \psi$, obtained from the expansion to quadratic order of (\ref{A3Fs}), contains both $\delta \psi''$ and $\delta \psi'$. One can transform it to Shrodinger form, i.e. remove the first derivative term, by a coordinate transformation $z \rightarrow \hat{z}$ followed by a $\hat{z}$-dependent field redefinition $\delta \psi \rightarrow \delta \hat{\psi}$. {}\footnote{In \cite{ASW2}, there was no need of a transformation $z \rightarrow \hat{z}$ since the coefficients of the $m^2$ and of the two-derivative terms had the same $z$-dependence. In more general examples, though, those coefficients can be different functions of $z$. In such cases, one needs the transformation $d\hat{z} = \sqrt{b(z)} \,dz$, where $b(z) = \frac{F_2 [F_1^{(1)} + F_1^{(2)} \psi'^2_{cl}]}{F_1^{(1)} F_1^{(2)}}$\,. Clearly, for this to be possible, $b(z)$ has to be positive-definite. It is, in fact, easy to verify that this is always the case by using the defining properties of the original background metric, namely that all spatial distance-squareds have to be positive-definite (which, for example, implies the positivity of the $zz$ metric component) and that all spatial subspaces have to have non-degenerate volume forms (which implies the positive-definiteness of the relevant subdeterminants).} For simplicity of notation, we will drop the hats in the following. 

The Schrodinger equation for the fluctuations has the form:
\be\label{eq}
-\delta\psi''(z)+V(z)\,\delta\psi(z)=m^2\,\delta\psi(z) \, ,
\ee
where we have suppressed the $x^{\mu}$ argument of $\delta \psi$ since it is only a spectator here. Let us recall a few, more or less well known, facts about the solutions of this equation. First, note that it is enough to study the range $z>0$, since the fluctuations can be split into symmetric and anti-symmetric ones w.r.t. to the point $z=0$. The range of $z$ can be either infinite or finite, in case a physical cutoff $z_\Lambda$ is imposed. The potential $V(z)$ is assumed to be bounded, except possibly at $z=0$. If the potential behaves as $|V(z)|\sim z^{-\lambda}$ with $\lambda<1$, then there are two complete sets of orthogonal solutions, one satisfying a Dirichlet boundary condition and the other satisfying a Neumann boundary condition. If the singularity of the potential is stronger, i.e. $2> \lambda\geq 1$, then one has only the complete system with Dirichlet boundary conditions. The other set of solutions is finite at $z=0$, but has a singular derivative at that point. Nevertheless, we will loosely refer to the latter solutions as satisfying Neumann boundary conditions even for $1\leq \lambda<2$, which are the only cases we will consider in this paper. The case $\lambda=2$ is more complicated and will not be discussed here. All of the statements in this paragraph can be read off from the exact solution of (\ref{eq}) for $m=0$ and $V(z)=-v_0 \,z^{-\lambda}$ with $v_0 = const >0 $, which is:
\be\label{solnlambda}
\psi_0(z)=\sqrt{z}\left[c_J\,J_{1\,/\,(2-\lambda)}\left(\frac{2\,z^{1-\lambda\,/\,2}\,\sqrt{v_0}}{2-\lambda}\right)+c_Y\,Y_{1\,/\,(2-\lambda)}\left(\frac{2\,z^{1-\lambda\,/\,2}\,\sqrt{v_0}}{2-\lambda}\right)\right].
\ee

\subsection{Positivity of ground state energy for low enough cutoff}

Let us consider (\ref{eq}), but with an auxiliary potential $V_{\zeta} (z)$, defined as $V_\zeta(z)=V(z)$ for $z<\zeta$ and $V_{\zeta}(z)=\infty$ for $z\geq \zeta$, instead of the potential $V(z)$. It is easy to realize that at sufficiently low cutoff $\zeta$ the ground state energy for this problem is positive, i.e. we have $m_0{}^2>0$. To show this, it is sufficient to consider attractive potentials, since a potential that is repulsive near $z=0$ always has a positive spectrum at small $\zeta$. 

For attractive potentials, the above statement ca be proven by scaling arguments. Indeed, such a potential can be replaced, at sufficiently low $\zeta$, by its asymptotic form
\be
V_\zeta(z)\simeq -\frac{v}{z^\lambda} \, ,
\ee
where the constant $v>0$. Then, if we rescale the coordinate by introducing $x=z\,/\,\zeta$ with the range of $x$ being $0\leq x \leq 1$, we arrive at the equation:
\be
-\delta\psi''(x)-\frac{v_0\,\zeta^{2-\lambda}}{x^\lambda}\,\psi(x)=(m\,\zeta)^2\,\psi(x).
\ee
Then, for $\lambda<2$, the potential term vanishes in the limit of $\zeta\to0$ and the spectrum reduces to that of a rectangular box potential with both the Dirichlet and Neumann spectra having positive eigenvalues $m^2$, which go to $+\infty$ as $\zeta^{-2}$.

\subsection{Dependence of ground state energy on cutoff value}

Let us denote by $\zeta_1$, $\zeta_2$ two cutoffs satisfying $\zeta_2>\zeta_1$. Then, clearly, in the interval $\zeta_1<z<\zeta_2$ we have $V_{\zeta_1}(z)=\infty>V_{\zeta_2}(z)$ and, furthermore, $V_{\zeta_1}(z)\geq V_{\zeta_2}(z)$ for every $z$. Let us also denote the ground state eigenfunctions by $\psi_{\zeta_1}$ and $\psi_{\zeta_2}$ and the ground state energy eigenvalues by $ m_{\zeta_1}{}^2$ and $ m_{\zeta_2}{}^2$. Since $\psi_{\zeta_1}(z)=0$ for $z>\zeta_1$, but  $\psi_{\zeta_1}(z)=\psi_{\zeta_2}(z)$ and $V_{\zeta_2}(z)=V_{\zeta_1}(z)$ for $z<\zeta_1$, we can write: 
 \be\label{extension}
 m_{\zeta_1}{}^2=\frac{\int_0^{\zeta_2
}dz\, \left\{\left[\psi_{\zeta_1}'\right]^2+V_{\zeta_2}(z)\,[\psi_{\zeta_1}]^2\right\}}{\int_0^{\zeta_2}dz\,[\psi_{\zeta_1}]^2} \, .
 \ee
Now, the right hand side of (\ref{extension}) can also be regarded as a variational estimate for $ m_{\zeta_2}{}^2$. However, this does not give the best estimate. Indeed, the right hand side of (\ref{extension}) is lowered upon replacing $\psi_{\zeta_1}$ by $\psi_{\zeta_2}$. Thus, the corresponding eigenvalues satisfy $ m_{\zeta_1}{}^2>m_{\zeta_2}{}^2$. This shows that the ground state eigenvalue decreases when the cutoff $\zeta$ is increased. In other words, the ground state energy is a monotonically decreasing function of the cutoff.

\subsection{Relation between zero-mass fluctuations and stability}

Let us now consider zero mass fluctuations, satisfying Dirichlet or Neumann boundary conditions at $z=0$. Then, according to Section 2.1, the ground state energy is positive at a sufficiently small cutoff $\zeta$. According to Section 2.2, the ground state energy decreases monotonically with increasing the cutoff. Hence, there are two possibilities: (1) the ground state energy stays positive up to a physical cutoff $z_\Lambda$ ($z_\Lambda$ can be finite or infinite), in which case there are no tachyons and the classical solution is perturbatively stable; (2) the ground state energy has a zero at a critical cutoff $z_c$. If $z_\Lambda > z_c$, then the ground state energy must be negative. The appearance of a tachyon in this spectrum signifies that the classical solution is perturbatively unstable. 

Let us denote by $\psi_0(z)$ a solution of (\ref{eq}) with $m=0$, satisfying either Dirichlet or Neumann boundary conditions at $z=0$. If $\psi_0(z_c)=0$ and  $\psi_0(z)\not=0$ for $z<z_c$, then at every physical cutoff $\zeta>z_c$ (including $\zeta=\infty$) the classical solution is unstable. On the other hand, for every cutoff, such that $\zeta\leq z_c$, the classical solution is stable.  In other words, the question of stability of classical solutions is reduced to the search for zeros of massless solutions of the equation (\ref{eq}), satisfying Dirichlet or Neumann boundary conditions.

\section{Massless fluctuations from classical solutions}
\setcounter{equation}{0}

Depending on the complexity of the potential $V(z)$, finding analytic solutions of (\ref{eq}) can be difficult, even if one considers solutions with $m=0$. In this section we will show that, once the general classical solution is found, finding the analytic expression for $m=0$ fluctuations around it is straightforward. 

As recalled in Section 2, the DBI action we study has the form:
\be\label{action}
S=\int dz\,d^4x\,\sqrt{F_1[\psi(z,x),\psi'(z,x),z]+F_2[\psi(z,x),z]\,\left[\partial_\mu\psi(z,x)\partial^\mu\psi(z,x)\right]} \, ,
\ee
where $x^{\mu}$ are the four-dimensional space-time coordinates. 
As explained in the previous section, the classical configurations of interest are the extrema of this action, which are independent of $x^{\mu}$. Clearly, when searching for them, one can drop the second term under the square root in (\ref{action}), thus reducing the problem to finding extrema of the action
\be\label{action2}
S=\int dz\,\sqrt{F_1[\psi(z),\psi'(z),z]} \, .
\ee

Now let us consider small fluctuations around a classical solution $\psi (z)$. For that purpose, we need to make the substitution
\be \label{fluctuations}
\psi(z,x) \to \psi(z)+\delta\psi(z,x)
\ee
in (\ref{action}). The next step is to find the linearized equation of motion for the fluctuation $\delta\psi(z,x)$. To do this, we expand the DBI action to second order in $\delta\psi(z,x)$. The only term of this effective action, which depends on derivatives with respect to $x^{\mu}$, is the kinetic term. Writing out the latter explicitly, the action has the form
\be\label{effaction}
S = \int\,dz\,d^4x\left[ \frac{F_2(\psi(z),z)\,\partial_\mu\psi(z,x)\partial^\mu\psi(z,x)}{\sqrt{F_1(\psi(z),\psi'(z),z)}}+...\right] \, ,
\ee
where the ellipsis refers to all the rest of the terms, other than the kinetic one. Expanding $\psi(z,x)$ in plane waves, we find that the kinetic term for a mass eigenstate reduces to: 
\be\label{kinterm}
S_K=m^2\int\,dz\,d^4x\, \frac{F_2(\psi(z),z)\,\psi^2(z)}{\sqrt{F_1(\psi(z),\psi'(z),z)}} \,\, .
\ee
Clearly, the action (\ref{kinterm}) vanishes for $m=0$. So we arrive at the (natural) conclusion that, at the linearized level, the massless fluctuations satisfy the same equation of motion as the classical solutions. Therefore, adding a small $m=0$ fluctuation to a classical solution results in another classical solution.  

Now, classical solutions satisfy second order differential equations and are, thus, parametrized by two integration constants. Hence, the only variations of a classical solution, that lead to another classical solution, are obtained by the variation of those integration constants. In particular, small variations of a classical solution correspond to infinitesimal changes of the integration constants. 

To write down the mathematical expression encoding the above statement, let us introduce some useful notation. First, we denote the two integration constants for classical solutions by $\lambda_{1,2}$. Also, to underline the dependence of such solutions $\psi_{cl} (z)$ on $\lambda_{1,2}$, let us write explicitly $\psi_{cl} (z,\lambda_1,\lambda_2)$. Then, the general solution of the linearized fluctuation equation for the $m=0$ case can be written as:
\be
\delta\psi_0(z,\lambda_1,\lambda_2)=c_1\, \partial_{\lambda_1}\,\psi(z,\lambda_1,\lambda_2)+c_2\, \partial_{\lambda_2}\,\psi(z,\lambda_1,\lambda_2) \, ,
\ee
where $c_1$ and $c_2$ are arbitrary constants.

Clearly, there are two linearly independent solutions $\delta\psi_0(z,\lambda_1,\lambda_2)$, of which, with an appropriate choice of $c_1$ and $c_2$, one combination satisfies Dirichlet and the other one satisfies Neumann boundary conditions at $z=0$. Let us denote those combinations by $\delta\psi^D_0(z,\lambda_1,\lambda_2)$ and $\delta\psi^N_0(z,\lambda_1,\lambda_2)$, respectively. According to the results of the previous section, a zero of these functions in the physical range of $z$ is a sign of instability of the classical solution $\psi_{cl} (z,\lambda_1,\lambda_2)$. Thus, to find whether a classical solution is stable or not, one only needs to investigate how it depends on its integration constants. In the next section, we will discuss applications of these results for several interesting examples.

\section{Examples}
\setcounter{equation}{0}

In this section, we will illustrate the great efficiency of the method, developed above, by applying it to several examples that are of significant phenomenological interest. We want to underline that the power of our method is in the following: To establish perturbative stability (or show perturbative instability) of a configuration of nonsupersymmetric probe branes, embedded in a nontrivial gravitational and flux background, one does not have to compute the whole scalar spectrum, arising from fluctuations of the embedding. Instead, it is enough to only investigate the massless modes in this spectrum. Furthermore, whenever the classical solution is known analytically, it is sufficient to study its derivatives with respect to the two integration constants it contains. The absence (presence) of zeros of those derivatives then signifies stability (instability) of the solution under consideration.

\subsection{D7-$\overline{{\rm D}7}$ branes in a deformed conifold background}
     
In \cite{KS}, Kuperstein and Sonneschein studied a model of flavor chiral symmetry breaking, obtained by embedding D7-$\overline{{\rm D}7}$ branes in a deformed conifold. The embedding is described by two functions, $\theta (r)$ and $\phi (r)$, of a radial variable $r$. The relevant DBI Lagrangian acquires the form:
\be\label{KSDBI}
L\sim r^3\,\left[1+\frac{r^2}{6}\left(\theta_r{}^2+\sin^2\,\theta\,\phi_r{}^2\right)\right]^{1/2} \, .
\ee

One can readily solve the Euler-Lagrange equation for the above Lagrangian. The solution is given by the classical configuration $\theta_{\rm cl}=\pi/2$ and
\be \label{phi_cl}
\phi_{\rm cl} (r)=\frac{\sqrt{6}}{4} \,\cos^{-1}\left[\left(\frac{r_0}{r}\right)^4\right] \, ,
\ee
where $r_0$ is an integration constant\footnote{As in \cite{KS}, we have chosen the value of $\phi$ at the tip to be zero.}. Hence, the results of the previous section imply the form
\be\label{deltaphi}
\delta\phi_0= c \,\, \partial_{r_0}\phi(r)\sim \frac{c}{\sqrt{r^8-r_0{}^8}}
\ee
for zero mass fluctuations of the field $\phi$, with $c$ being a constant. Clearly, (\ref{deltaphi}) is singular at $r=r_0$, i.e. at the point where the D7 and $\overline{{\rm D}7}$ branes merge and where one has to impose boundary conditions. This signifies the need of a change of coordinates, in order to properly describe the physics of the fluctuations around the classical solution.
   
A suitable choice are the Cartesian coordinates $y$ and $z$ defined as \cite{KS}:
\bea
y&=&r^4\,\cos(4\,\phi\,/\,\sqrt{6}) \, ,\nn\\
z&=&r^4\,\sin(4\,\phi\,/\,\sqrt{6}) \, .
\eea
It is easy to see that, in terms of the variable $y$, the classical solution (\ref{phi_cl}) acquires the form: $y_{\rm cl}=r_0^4$. Hence any fluctuation of $y$ is transverse to the D7-$\overline{{\rm D}7}$ embedding. On the other hand, $z$ is a worldvolume coordinate. So, in order to study transverse fluctuations, we consider $y$ and $\theta$ as functions of $z$. The relevant DBI action has the form:
\be
L\sim \frac{\sqrt{3\,\sin^2\theta\,(y-z\,y')^2+3\,(z+y\,y')^2+8\,(z^2+y^2)^2\,\theta'{}^2}}{\sqrt{z^2+y^2}} \, .
\ee
   
Substituting the classical value $\theta=\pi\,/\,2$, the Lagrangian reduces to:
\be
L_{\pi/2}\sim \sqrt{1+y'{}^2}.
\ee
Then the solution of the equation of motion for $y(z)$ is 
\be\label{classicaly}
y(z)= a+b\,z \, ,
\ee
where $a$ and $b$ are integration constants. The derivatives of $y(z)$ with respect to $a$ and $b$ provide the two zero mass fluctuations $\delta y_1=1$ and $\delta y_2=z$, which satisfy Neumann and Dirichlet boundary conditions at $z=0$, respectively. These fluctuations do not vanish anywhere in the physical range of $z$, $-\infty<z<\infty$ (except for the trivial zero of the Dirichlet solution). Hence, the classical solution is stable with respect to fluctuations in the $y$ direction.
        
Finding the massless fluctuations of $\theta$ is slightly more complicated. The reason is that we have not been able to find an analytic expression for the general solution of the Euler-Lagrange equation for this field. Nevertheless, we can investigate zero mass fluctuations $\delta \theta$ around the classical solution $\theta_{cl}=\pi/2$. Using that $y_{cl}=r_0^4$, we obtain the following linearized equation of motion for $\delta\theta$:
\be\label{deltathetalinear}
(r_0{}^8+z^2)\,\delta\theta''=2\,z\,\delta\theta'+\frac{3\,r_0{}^8}{r_0{}^8+z^2}\,\delta\theta.
\ee
The general solution of (\ref{deltathetalinear}) is:
\be
\delta\theta=c_D \,\sin\left[\frac{\sqrt{3}}{2\,\sqrt{2}}\,\tan^{-1}(z\,/\,r_0{}^4)\right]+c_N \,\cos\left[\frac{\sqrt{3}}{2\,\sqrt{2}}\,\tan^{-1}(z\,/\,r_0{}^4)\right] \, ,
\ee
where the terms with $c_D$ and $c_N$ satisfy Dirichlet and Neumann and boundary conditions at $z=0$ respectively. Clearly, this solution is regular and does not vanish at any $z\neq 0$ point. Thus, the D7-$\overline{{\rm D}7}$ embedding under consideration is perturbatively stable. Of course, this result was obtained via direct mass-spectrum calculations in \cite{KS}. Nevertheless, the new derivation presented here illustrates the power of our method.

\subsection{Sakai-Sugimoto model: D8-$\overline{{\rm D}8}$ probes in D4 background}

The second example we consider is the Sakai-Sugimoto holographic dual of large $N_c$ QCD \cite{SS}. This model is based on a U-shaped D8-$\overline{{\rm D}8}$ flavor branes embedding in a D4 brane background. Now there is only one direction, transverse to the flavor brane probes, namely a sphere parametrized by a coordinate $\tau$. The radial coordinate is denoted by $U$ \cite{SS}. So the D8-$\overline{{\rm D}8}$ position in the transverse space is described by a function $\tau (U)$.

The classical solution for the embedding function $\tau(U)$ can be obtained easily by noticing that
the DBI Lagrangian does not depend explicitly on the variable $\tau$. Hence, the system is conservative and the classical solution can be obtained from the first integral $\partial_\tau \,H=0$. The result is:
\be\label{classical}
\tau(U)=\tau_0+U_0{}^4\,f(U_0)^{1/2}\,\int_{U_0}^U\left(\frac{R}{U}\right)^{3/2}\,\frac{dU}{f(U)\,\sqrt{U^8\,f(U)-U_0{}^8\,f(U_0)}} \, ,
\ee
where $f(U)=1- U_{KK}{}^3\,/\,U^3$ and $U_0>U_{KK}$. While $U_{KK}$ is a physical  scale parameter, $\tau_0$ and $U_0\geq U_{KK}$ are integration constants. If one choses $\tau_0=0$, then the initial condition is $\tau(U_0)=0$. The work \cite{SS} considered the case $U_0=U_{KK}$, while \cite{CES} studied $U_0 > U_{KK}$ albeit in a different phenomenological application of that model. Those papers showed perturbative stability by computing explicitly the mass spectrum, thus proving that it does not contain tachyonic modes.\footnote{In fact, the work \cite{CES} studied only the (axial-)vector spectrum, but not the scalar one. Hence, the question of stability of the $U_0 > U_{KK}$ case was not settled.}

Now we will recover the same result by applying the new method developed here. According to the previous two sections, to find whether the classical solution is stable or not, we need to calculate its derivatives with respect to the two integration constants $\tau_0$ and $U_0$. The derivative with respect to $\tau_0$ is the fluctuation $\delta\tau_N(U)=1$, which satisfies a Neumann boundary condition at $U=U_0$. Obviously, it does not have zeros as a function of $U$. 
Finding the derivative with respect to $U_0$ is more demanding. For a generic choice of $U_0$, \cite{SS} only showed that $d\tau\,/\,dU_0<0$ in the limit $U\to\infty$. This does not rule out an instability, if the sign of the derivative were to change at a smaller value of $U$. We will investigate this issue with our new method.

Unfortunately, we cannot write down the general analytic form of $\pd_{U_0}\tau(U)$. But we will be able to analyze explicitly two special cases. The first is when one chooses the integration constant $U_0$ such that $U_0>\!\!>U_{KK}$. In this case, one can approximate $f(U)\simeq1$. Then one can evaluate the integral in (\ref{classical}) in terms of a hypergeometric function. Calculating the derivative with respect to $U_0$ and using for convenience the rescaled variable $V=U\,/\,U_0$, we find:
\be
\delta\tau(U)=\partial_{U_0}\tau(U)\sim  \frac{1}{V^{9/2}}\,{}_2F_1\left(\frac{1}{2},\frac{9}{16};\frac{25}{16};\frac{1}{V^{8}}\right)-\frac{9}{\,\sqrt{V(V^8-1)}}-\frac{9\,\sqrt{\pi}\,\,\Gamma(9\,/\,16)}{\Gamma(1\,/\,16)} \, .
\ee
This expression is dominated by the manifestly negative second and third terms at $V\to1$ and $V\to\infty$, respectively. The derivative of this expression with respect to $U$ is $36 \, U^{13/2}\,(U^8-1)^{3/2}>0$. Therefore, $\delta\tau(U)$ is a monotonically rising function and is negative everywhere. Consequently, the classical solution (\ref{classical}) for the Sakai-Sugimoto  D4\,/\,D8/\,$\overline{{\rm D}8}$ system is stable for $U_0>\!\!>U_{KK}$.

Another special case, that is tractable analytically, is given by choosing $U_0$ in a small neighborhood of $U_{KK}$. In other words, we take the integration constant to be such that $U_0-U_{KK}<\!\!<U_{KK}$. Notice that the integral in (\ref{classical}) is divergent for $U_0=U_{KK}$. Nevertheless, it is possible to determine the sign of the leading term. 
As a first step, we integrate (\ref{classical}) by parts, in order to make the integral non-singular at the lower limit. This gives:
\be\label{classical2}
\tau(U)\sim U_0{}^4\,f(U_0)^{1/2}\,\int_{U_0}^U dU\,\frac{(136\,U^6+25\,U_{KK}^6-143\,U^3\,U_{KK}^3)\sqrt{U^8\,f(U)-U_0{}^8\,f(U_0)}}{U^{7/2}\,(5\,U_{KK}{}^6+8\,U^6-13\,U^3\,U_{KK}^3)^2} \, .
\ee
The derivative of the integral in (\ref{classical2}) with respect to $U_0$ is singular for $U_0\to U_{KK}$. The singular behavior comes from the region in which $U-U_0={\cal O}(U_0-U_{KK})$. Therefore, to calculate the leading order contribution, it is sufficient to take the derivatives of the singular multipliers, such as $f(U_0)^{1/2}$ and $\sqrt{U^8\,f(U)-U_0{}^8\,f(U_0)}$, and then to substitute $U\to U_{KK}$ in the non-singular multipliers. As a result, we obtain:
\be
\partial_{U_0}\tau(U)\sim \int_{U_0}^U dU \frac{\partial_{U_0}\left[f(U_0)^{1/2}\,\sqrt{U^8\,f(U)-U_0{}^8\,f(U_0)}\right]}{\left(5\,U_{KK}{}^6+8\,U^6-13\,U^3\,U_{KK}^3\right)^2} \, .
\ee
Let us now introduce the small parameter $\Delta=(U_0-U_{KK})\,/\,U_{KK}$. Also, it will be convenient to change variables $U \to u$ via $U=U_{KK}[1+\Delta(1+u)]$. Then, substituting  $U_0=(1+\Delta)\,U_{KK}$, we find that the leading order contribution in a power series in $\Delta$ is:
\be
\partial_{U_0}\tau(U)\sim -\frac{1}{\Delta}\int_0^u du\frac{1-u}{(1+u)^2\sqrt{u}}=-\frac{2}{\Delta}\frac{\sqrt{u}}{1+u}\sim -\frac{1}{U_0-U_{KK}}\frac{\sqrt{U-U_{0}}}{U-U_{KK}}<0 \, .
\ee
This implies that $\delta\tau(U)<0$ for all $U>U_0$, which proves the stability of the classical solution for the D8-$\overline{{\rm D}8}$ embedding in the parameter space region $U_0-U_{KK}<\!\!<U_{KK}$ as well.

\subsection{Walking technicolor model: D7-$\overline{{\rm D}7}$ branes in D5 background} 

Another interesting example is the model of dynamical electroweak symmetry breaking, obtained by embedding D7-$\overline{{\rm D}7}$ probes \cite{LA} in the gravitational background of \cite{NPP,EGNP} that is dual to a walking gauge theory. In \cite{ASW3}, we applied a part of our method to this model. This application provides a useful complement to the previous two examples and so we will briefly review it here. 

The relevant DBI Lagrangian takes the form:
\be\label{dbi}
L_{DBI}\sim e^{2\,\rho}\sqrt{\frac{4}{3}\,\beta\,e^{4\,\rho}+\theta_\rho{}^2+\sin^2(\theta)\,\varphi_\rho{}^2} \, ,
\ee
where $\beta$ is a small parameter characterizing the length of the energy region in which the gauge coupling is walking. The variable $\rho \ge 0$ parameterizes the radial direction and is larger than 1 in the walking region. The embedding in the transverse space is described by the angular variables $\theta$ and $\varphi$ as functions of $\rho$; we have denoted $\frac{d \theta}{d \rho} \equiv \theta_\rho$ and $\frac{d \varphi}{d \rho} \equiv \varphi_\rho$. It is easy to see that the constant choice $\theta=\frac{\pi}{2}$ satisfies the equation of motion for the field $\theta (\rho)$. It is also easy to show that fluctuations $\delta \theta$ around this classical solution are stable.\footnote{To second order, the fluctuations of $\theta$ and $\varphi$ decouple.} Therefore, we will concentrate on studying the field $\varphi (\rho)$, while setting $\theta = \frac{\pi}{2}$.

The classical solution for $\varphi (\rho)$ is a nontrivial function. Due to that, the transverse direction with respect to the D7-$\overline{{\rm D}7}$ embedding is a combination of both $\varphi$ and $\rho$. Hence, to capture the full fluctuation spectrum, one should change coordinates, as explained in \cite{ASW3}. A suitable choice is the following \cite{ASW3}:
\bea
y&=& \frac{e^{2\,\rho}}{\cosh\left(\psi\right)} \, ,\nn\\
z&=& e^{2\,\rho}\,\tanh\left(\psi\right) \, ,
\eea
where we defined $\psi=\varphi\,e^{-2\,\rho_0}\,\sqrt{3\,/\,\beta}$ for convenience. Note that these coordinates satisfy the Cartesian-like relation
\be
y^2+z^2 = e^{4\,\rho} \, .
\ee
Now the classical solution is given by $y_{cl}=e^{2 \,\rho_0}$ with $z$ arbitrary; see \cite{ASW3}. Clearly then, $z$ runs along the classical embedding, whereas fluctuations of $y$ are transverse to it. Hence, to study the transverse fluctuations, we need to consider the field $y(z)$ and expand it as $y(z) = y_{cl} + \delta y(z)$.
      
The linearized equation of motion for massless $\delta y (z)$ fluctuations acquires the form:
\be \label{leom}
(e^{4\,\rho_0}+z^2)^2\,\delta y''+z\, (e^{4\,\rho_0}+z^2)\,\delta y'+2\,e^{4\,\rho_0}\,\delta y=0.
\ee
The general solution of (\ref{leom}) is:
\be \label{yfluc}
\delta y= c_D \,\frac{z}{\sqrt{z^2+e^{4\,\rho_0}}}+c_N\,\left[-1 +\frac{z}{\sqrt{z^2+e^{4\,\rho_0}}}\log\left(e^{-2\,\rho_0}(z+\sqrt{z^2+e^{4\,\rho_0}})\right)\right] \, ,
\ee
where the coefficients of $c_D$ and $c_N$ satisfy Dirichlet and Neumann boundary conditions at $z=0$, respectively. The second term vanishes at the point $z=e^{2\,\rho_0}\,z_c$, where we have denoted by $z_c\simeq 1.5$ a certain critical value. Therefore, according to the results of the previous sections, for sufficiently large cutoff $z_\Lambda >e^{2\,\rho_0}\,z_c$ the spectrum of excitations contains a tachyon and the system is unstable.

We should note that an instability in this model was first indicated in \cite{CLtV}, in the context of evaluating numerically the mass spectrum of scalar fluctuations.\footnote{For a more detailed comparison between \cite{CLtV} and \cite{ASW2,ASW3}, see \cite{ASW3}. We are grateful to T. ter Veldhuis for an extensive discussion regarding the work of \cite{CLtV}.} Our result for $z_c$, which is determined from the single root of the Neumann term of (\ref{yfluc}), agrees with their numerical estimate for the critical value of the cutoff.

\section{Summary}

We developed a simple method for the investigation of perturbative stability of nonsupersymmetric  probe branes embedded in nontrivial gravitational and flux backgrounds. The method relies on the key statement that zeros of massless fluctuations, around a given classical configuration, indicate the presence of an instability (in the form of a tachyonic mode in the fluctuation spectrum). Therefore, it is enough to investigate the behavior of the zero mass fluctuations. Furthermore, 
the analytic form of the latter can be derived directly from the classical solution under consideration, by taking derivatives with respect to its integration constants. 

We illustrated our method by applying it to three examples of phenomenological interest: the model of chiral symmetry breaking of \cite{KS}, the holographic dual of strongly coupled large $N_c$ QCD proposed in \cite{SS}, as well as its modification considered in \cite{CES}, and the dynamical electroweak symmetry breaking model of \cite{LA,ASW}. In all of these cases, we recovered, and in the case of \cite{CES} extended, the previously known results regarding the issue of stability. However, we did not need to compute the full scalar mass spectrum. This is rather important, as quite often the computation of this spectrum is only feasible by numerical methods. 

Work on other, more systematic applications of our method is in progress.

\section*{Acknowledgements}

We would like to thank T. ter Veldhuis for correspondence regarding Ref. \cite{CLtV} and P. Argyres for valuable discussions. The research of P.S. and R.W. has been supported by DOE grant FG02-84-ER40153. Research at Perimeter Institute is supported by the Government of Canada through Industry Canada and by the Province of Ontario through the Ministry of Research \& Innovation.

 \end{document}